# Search for Low-Mass Exoplanets by Gravitational Microlensing at High Magnification

F. Abe,[1] D.P. Bennett,[2] I.A. Bond,[3] S. Eguchi,[1] Y. Furuta,[1] J.B. Hearnshaw,[4] K. Kamiya,[1] P.M. Kilmartin,[4] Y. Kurata,[1] K. Masuda,[1] Y. Matsubara,[1] Y. Muraki,[1] S. Noda,[5] K. Okajima,[1] A. Rakich,[6] N.J. Rattenbury,[7*] T. Sako,[1] T. Sekiguchi,[1] D.J. Sullivan,[8] T. Sumi,[9] P.J. Tristram,[10] T. Yanagisawa,[11] P.C.M. Yock,[10] A. Gal-Yam,[12,13] Y. Lipkin,[14] D. Maoz,[14] E.O. Ofek,[14] A. Udalski,[15] O. Szewczyk,[15] K. Żebruń,[15] I. Soszyński,[15] M.K. Szymański,[15] M. Kubiak,[15] G. Pietrzyński,[15,16] L. Wyrzykowski[15]

[1]Solar Terrestrial Environment Laboratory, Nagoya University, Nagoya 464-01, Japan. [2]Department of Physics, Notre Dame University, Notre Dame IN 46556, USA. [3]Institute of Information & Mathematical Sciences, Massey University at Albany, Auckland, NZ. [4]Department of Physcs and Astronomy, University of Canterbury, Christchurch, NZ. [5]National Astronomical Observatory, Tokyo, Japan. [6]Electro Optics Systems, Canberra, Australia. [7]Department of Physics and Astronomy, University of Manchester, Manchester, UK. [8]School of Chemical & Physical Sciences, Victoria University, Wellington, NZ. [9]Department of Astrophysical Sciences, Princeton University, Princeton NJ 08544, USA. [10]Department of Physics, University of Auckland, Auckland, NZ. [11]National Aerospace Laboratory, Tokyo, Japan. [12]Department of Astronomy, California Institute of Technology, Pasadena, CA 91025, USA. [13]Hubble Fellow. [14]School of Physics and Astronomy, Raymond and Beverley Sackler Faculty of Exact Sciences, Tel-Aviv University, Tel Aviv 69978, Israel. [15]Warsaw University Observatory, Al. Ujazdowskie 4, 00-478 Warszawa, Poland. [16]Universidad de Concepcion, Departamento de Fisica, Casilla 160-C, Concepcion, Chile.

* To whom correspondence should be addressed. E-mail: njr@jb.man.ac.uk

## ABSTRACT

Observations of the gravitational microlensing event MOA 2003-BLG-32/OGLE 2003-BLG-219 are presented for which the peak magnification was over 500, the highest yet reported. Continuous observations around the peak enabled a sensitive search for planets orbiting the lens star. No planets were detected. Planets 1.3 times heavier than Earth were excluded from more than 50 % of the projected annular region from approximately 2.3 to 3.6 astronomical units surrounding the lens star, Uranus-mass planets from 0.9 to 8.7 astronomical units, and planets 1.3 times heavier than Saturn from 0.2 to 60 astronomical units. These are the largest regions of sensitivity yet achieved in searches for extrasolar planets orbiting any star.

Gravitational microlensing events of high magnification occur when the foreground lens system comes into near-perfect alignment with the background source star. Suitable alignments are most readily found in the dense stellar fields in the Galactic bulge, where magnifications as high as 1000 are possible (*1*). In these events, the two images of the source star produced by the lens star merge to form a near annular single ring image. The events provide enhanced sensitivity to planetary companions of the lens star because they can, depending on the planetary mass and position, perturb the ring-like image of the source star at times near the peak amplification (*1-5*). They complement events of low magnification that also provide substantial sensitivity to extrasolar planets when 'caustic crossings' occur (*6-9*). For both detection methods, the sensitivity to low-mass planets is enhanced in events with small (that is, main-sequence) source stars (*2,5,8*).

High magnification events can be detected in microlensing surveys that are sensitive to unresolved main-sequence stars and employ frequent sampling of target fields (*10*). However, most events occur on very faint sources and, while spectacular, are effectively very short-lived. Here we report observations of the very high magnification microlensing event MOA 2003-BLG-32/OGLE 2003-BLG-219 (hereinafter M32/O219). It was the first high magnification event (with a non-binary lens star) for which the FWHM of the peak was intensively monitored. The observations were carried out with the 0.6-m telescope operated by the Microlensing Observations in Astrophysics (MOA) collaboration at the Mt John Observatory in New Zealand, the 1.0-m telescope of the Wise observatory in Israel, and the 1.3-m Warsaw telescope operated by the Optical Gravitational Lensing Experiment (OGLE) collaboration at the Las Campanas Observatory in Chile.

M32/O219 was found independently by the MOA and OGLE microlensing surveys (*11,12*). The event brightened rapidly in the MOA survey on 12 June 2003 at $\alpha$ = 18h:05m:34s.64 and $\delta$ = -29°:06':53".6 (J2000.0), i.e., at Galactic coordinates are $l$ = 2.04° and $b$ = -3.86°. Intensive observations were carried out by MOA, and subsequently by the Wise observatory during daylight hours in New Zealand. Unfortunately, the OGLE telescope was clouded out at the time, but constraining observations were made before the event peaked, and further constraining observations were made as it faded. The total dataset presented here comprise 44 Wise I-band images, 2 Wise V-band images, 921 MOA images from 2000-2003 in a wide red band, and 182 OGLE I-band images from 2001-2003. Except for the Wise V-band images, all images were analyzed using the difference-imaging technique to achieve the best possible photometric accuracy (*13*). This resulted in three independent sets of uncalibrated "delta-flux" measurements (tables at MOA website www.physics.auckland.ac.nz/moa/index.html).

The delta-flux measurements are well fitted by the theoretical light curve for microlensing of a point source by a single lens (Fig. 1). The light curve may be characterized by the normal parameters for gravitational microlensing. These are the Einstein radius crossing time $t_E$, the impact parameter $u_0$ of the source star trajectory with respect to the lens star in units of the Einstein radius $r_E$, and the time $t_0$ of closest approach of the source star to the lens star. The fitted values are $u_0$ = 0.00191 ± 0.00028, $t_0$ = HJD2452803.4856 ± 0.0001, and $t_E$ = (20.87 ± 0.11 day) × (0.00191/$u_0$) (*14*). The $\chi^2$ for the best fit has $\chi^2$/(degrees of freedom) = 1146.6 / 1129. The peak amplification, $A_{max}$, was 520 ± 80. Deviations of the light curve from that of a point source were searched for, but none were found. A 2σ upper limit of ~ 0.0016 for the ratio $r_s/r_{E'}$ was obtained, where $r_s$ denotes the radius of the source star, and $r_{E'}$ the Einstein radius projected to the plane of the source star.

The Wise images of the event were calibrated using the OGLE catalogue of bulge stars (*15*) and the DoPHOT photometry program (*16*) for crowded fields (tables at MOA website). This yielded model-independent values of the peak magnitude of the event, $I_{peak}$ = 14.251 ± 0.018, and of the color index of the source star, V – I = 1.51 ± 0.03. Here the small uncertainties result from the high magnification of the event. The microlensing fit yielded the baseline magnitude of the source star, I = 21.05 ± 0.15. This enabled the source star to be placed on a color-magnitude diagram of a nearby field (Baade's window) that was obtained by the Hubble Space Telescope (*17*). When allowance was made for the slightly different extinctions towards Baade's window and M32/O219, the source star for M32/O219 was found to lie on the perimeter of the HST diagram that corresponds to a metal-poor G-type main sequence star with radius ~ 0.7 R located at the back of the Galactic bulge (Fig. 2). The low metallicity of the source star was not anticipated, but we note that a similar result was reported recently for another event with a faint source star at the back of the Galactic bulge (*18*).

The sensitivity of high magnification microlensing events to the presence of extrasolar planets depends not only on the peak magnification, but also on the size of the source star, with smaller source stars providing higher sensitivity (*2,5,10*). The source-size enters through the ratio $r_s/r_{E'}$, which, as noted above, is restricted by the light curve of M32/O219 to values < 0.0016. For a lens of mass ~ 0.4 solar mass [typical for events with $t_E$ ~ 20 day (*20*)] located near the back of the bulge, $r_s/r_{E'}$ ~ 0.0015. For a lens of the same mass, but located towards the front of the bulge, $r_s/r_{E'}$ ~ 0.0007. The latter location is more likely, as it doubles the projected Einstein radius and the probability for lensing, and is entirely consistent with the absence of source-size effects in the light curve. In what follows, we report the sensitivity of M32/O219 to planets for $r_s/r_{E'}$ equal to both 0.0007 and 0.0015. Values lower than 0.0007 are also possible, but they correspond to less likely masses and/or locations of the lens star.

We note that $r_E$ ≈ 2.9 astronomical units (AU) if $r_s/r_{E'}$ = 0.0007. We also note that all the parameters in microlensing events, including $r_s/r_{E'}$, may be measured by observing the lens and the source as they diverge from one another after microlensing ceases (*21*), and that Monte Carlo simulations may also be carried out to constrain these parameters (*18*). Post-event observations would be especially valuable for those events in which planets are detected.

As the light curve of M32/O219 (Fig. 1) is consistent with that of a single lens, we used the data to determine exclusion zones for extrasolar planets orbiting the lens. Light curves were computed corresponding to the lens star with trial planets placed at various positions and with various masses, and compared to the observed data. The projected co-ordinates of the planet in the lensplane, $x_p$ and $y_p$, were both allowed to range from $-8r_E$ to $+8r_E$ in steps of $r_E/32$ for planet:star mass ratio values $q = 10^{-6}$ and $10^{-5}$, and from $-64r_E$ to $+64r_E$ in steps of $r_E/4$ for $q = 10^{-4}$ and $10^{-3}$. In a previous search for extrasolar planets by microlensing, a finer grid was used (*22*).

A numerical light curve for each trial planet was generated using an inverse ray-shooting algorithm (*5*), and the $\chi^2$ with respect to the observed data was calculated. The value of $\chi^2$ for the best fit curve without an extrasolar planet was subtracted to provide a difference of chi-squares, $\Delta\chi^2$, for each trial. A threshold value of the difference was set at $\Delta\chi^2 = 40$ as the minimum required to exclude the presence of an extrasolar planet. This is smaller than the threshold value used in (*22*). However, in the previous search, extrasolar planets were sought in events of lower magnification, where it is known that planetary perturbations may occur anywhere on the light curve during the Einstein crossing time, $2t_E$. In contrast, in events of high magnification like the present one, most planetary perturbations are confined to a short time window of order the FWHM of the light curve (*5*). This reduces the number of degrees of freedom, and permits a lower threshold value of $\Delta\chi^2$ to be used. Also, the photometry for the present study was carried out using the difference imaging technique (*13*). This utilizes all the information contained in the images of dense stellar fields to minimize systematic and statistical errors. Fig. 3 shows light curves for various trials with $\Delta\chi^2$ at the threshold value superimposed upon the data (*23*). Significant inconsistencies between these light curves and the data are seen.

Fig. 4 shows exclusion regions for planets as a function of the planet:star mass ratio, *q*, and the projected separation, *a*, from the lens star of M32/O219. These are the regions for which $\Delta\chi^2 >$ the threshold value given above. Fig. 5 shows planetary detection efficiencies obtained by integrating the exclusion regions over the possible position angles (0° - 360°) of extrasolar planets. The efficiencies exceed 50% for $a = (0.80 - 1.25)r_E$ for $q = 10^{-5}$, and for $a = (0.32 - 3.0)r_E$ and $(0.05 - 21)r_E$ for $q = 10^{-4}$ and $10^{-3}$ respectively. With the estimates given above, the $q$ values $10^{-6}$, $10^{-5}$, $10^{-4}$ and $10^{-3}$ correspond approximately to extrasolar planets of 1.3 times Mars mass, 1.3 times Earth-mass, 1.0 times Uranus mass and 1.3 times Saturn-mass, respectively, and, as noted above, $r_E \approx 2.9$ AU. The observations therefore imply that planets slightly heavier than Earth are excluded from more than 50% of the projected annular region from approximately 2.3 to 3.6 AU surrounding the lens star of M32/O219, Uranus-mass planets from 0.9 to 8.7 AU, and planets slightly heavier than Saturn from 0.2 to 60 AU.

Although no extrasolar planets were found in M32/O219, the observations eliminated the possible presence of terrestrial, ice-giant and gas-giant planets over large ranges of orbital radii. If several microlensing events of high magnification can be detected and monitored, upper limits or rough statistics will be able to be determined on the abundances of these planets. A few events with magnification exceeding 100 are detected annually (*10,11*), and events with the very high magnification of M32/O219 may not be uncommon (*18*). As our present knowledge of extrasolar planets is largely theoretical (e.g., *25-29*), any observational information should be helpful. Such information could also assist the planning of future, space-based probes of extrasolar planets (*30-32*).

33. The authors thank A. Gould for helpful comments and J. Holtzman for providing the HST data for Fig. 2. The MOA project is supported by the Marsden Fund of New Zealand, the Ministry of Education, Culture, Sports, Science, and Technology (MEXT) of Japan, and the Japan Society for the Promotion of Science (JSPS). Gravitational lensing studies at Wise Observatory are supported by a grant from the German Israeli Foundation for Scientific Research and Development. The OGLE group is supported by the following grants: Polish KBN 2P03D02124, "Subsydium Profesorskie" of the Foundation for Polish Science, NSF grant AST-0204908 and NASA grant NAG5-12212. D.B. acknowledges support from grants NSF AST-0206187 and NASA NAG5-13042. A.G. acknowledges support by NASA through Hubble Fellowship grant HST-HF-01158.01-A awarded by STScI, which is operated by AURA, Inc., for NASA, under contract NAS5-26555.

25 May 2004; accepted 28 July 2004

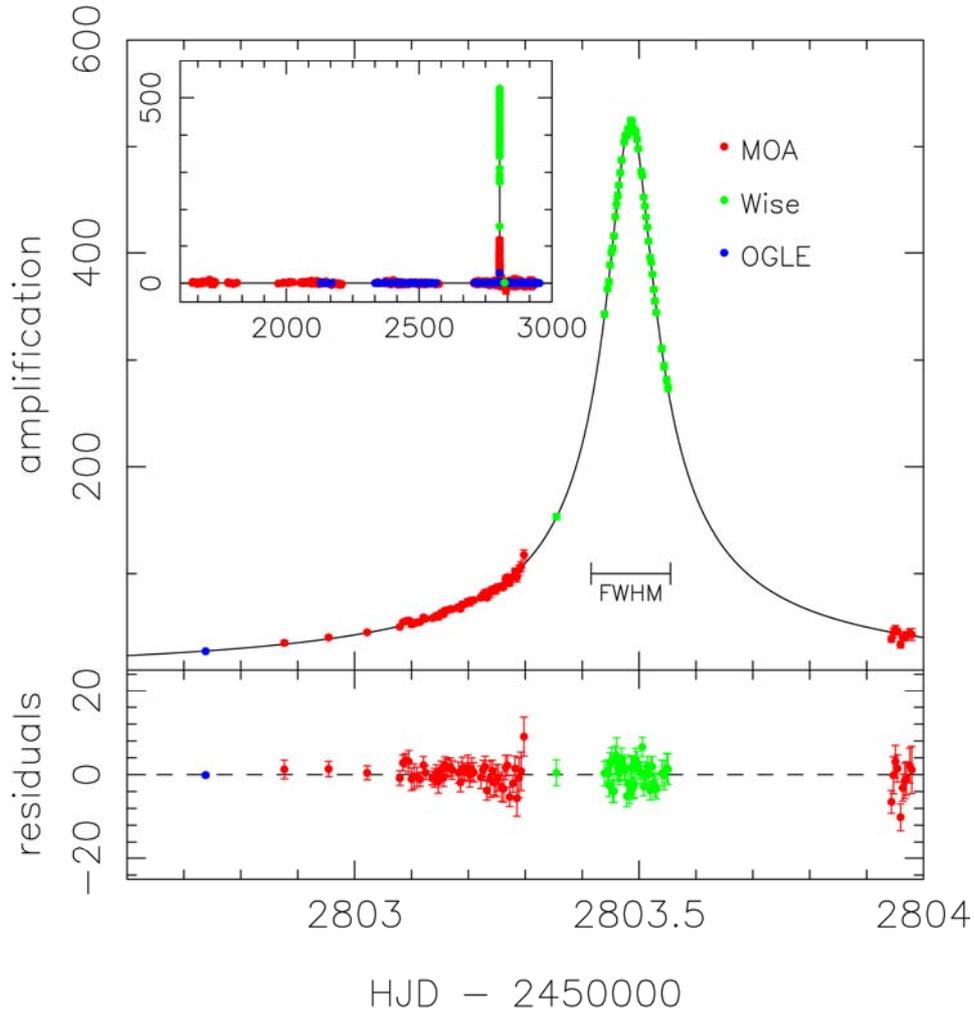

**Fig. 1.** Light curve of MOA 2003-BLG-32 obtained by MOA (red), OGLE (blue) and Wise (green) over 1.5 days and 4 years. The delta-fluxes obtained from the difference imaging analysis were normalized to amplifications that correspond to the best fit for a single lens and a point source, as shown by the solid line. HJD is heliocentric Julian day.

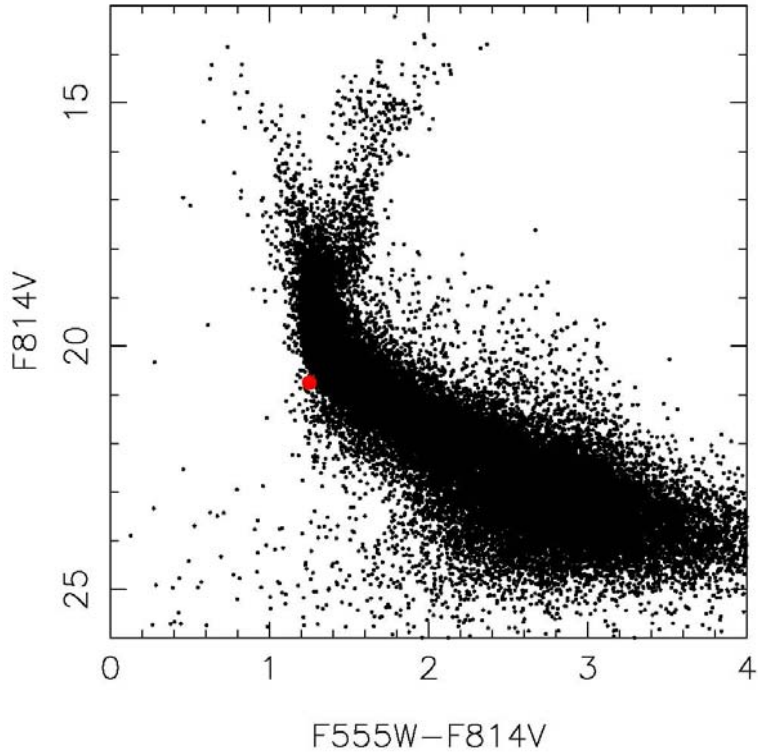

**Fig. 2.** Color-magnitude diagram obtained by the HST of Galactic bulge stars in Baade's window (*17*). The HST passbands F814V and F555W are similar to the conventional I and V passbands. The red dot denotes the position of the source star for M32/O219, after a small correction has been made for the slightly different values of extinction and reddening towards Baade's window and M32/O219. The position corresponds to the source star being located at the very back of the Galactic bulge. The de-reddened magnitude and color of the source star, determined from the position of the red-giant clump as in (*18*), are $I_0 = 19.9 \pm 0.2$ and $(V-I)_0 = 0.63 \pm 0.05$ respectively. Assuming a distance of $10.5 \pm 0.5$ kpc, the absolute magnitude is $M_I = 4.8 \pm 0.2$. The magnitude and color correspond to a low-metallicity G-type star of age ~ 10 Gy and radius ~ 0.7 R (*19*).

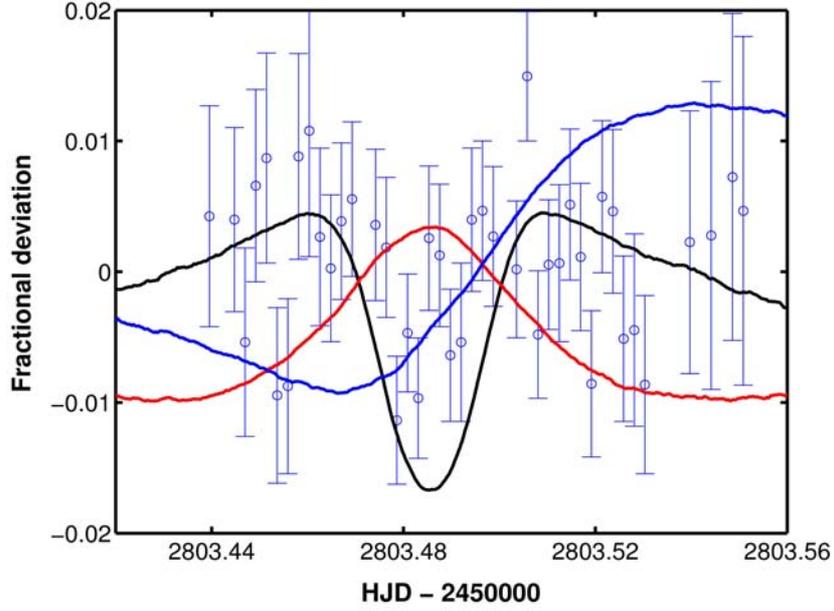

**Fig. 3.** Fractional deviations from the single lens for various trial planets with $\Delta\chi^2$ at the threshold value required for exclusion, corresponding to planets on the perimeters of the exclusion regions in Fig. 2. The mass fractions, $q$, and position angles of the planets as defined in (5), are: $10^{-5}$, 90°, black; $10^{-4}$, 270°, red; $10^{-3}$, 45°, blue. The source-size parameter $r_s/r_{E'} = 0.0007$. Discrepancies between the trials and the data are apparent. For planets in the interiors of the exclusion regions in Fig. 4, the discrepancies are larger, with $\Delta\chi^2$ reaching values as high as 4,800, 72,000 and 133,000 for $q = 10^{-5}$, $10^{-4}$ and $10^{-3}$ near the Einstein ring. The parameters $t_0$, $t_E$ and $u_0$ were held fixed at their values determined in the single lens fitting, with $u_0 = 0.00194$ for $r_s/r_{E'} = 0.0007$. Limb darkening of the source star was included following Claret (*24*). The fine-scale structure in the plots is caused by numerical noise and is unphysical.

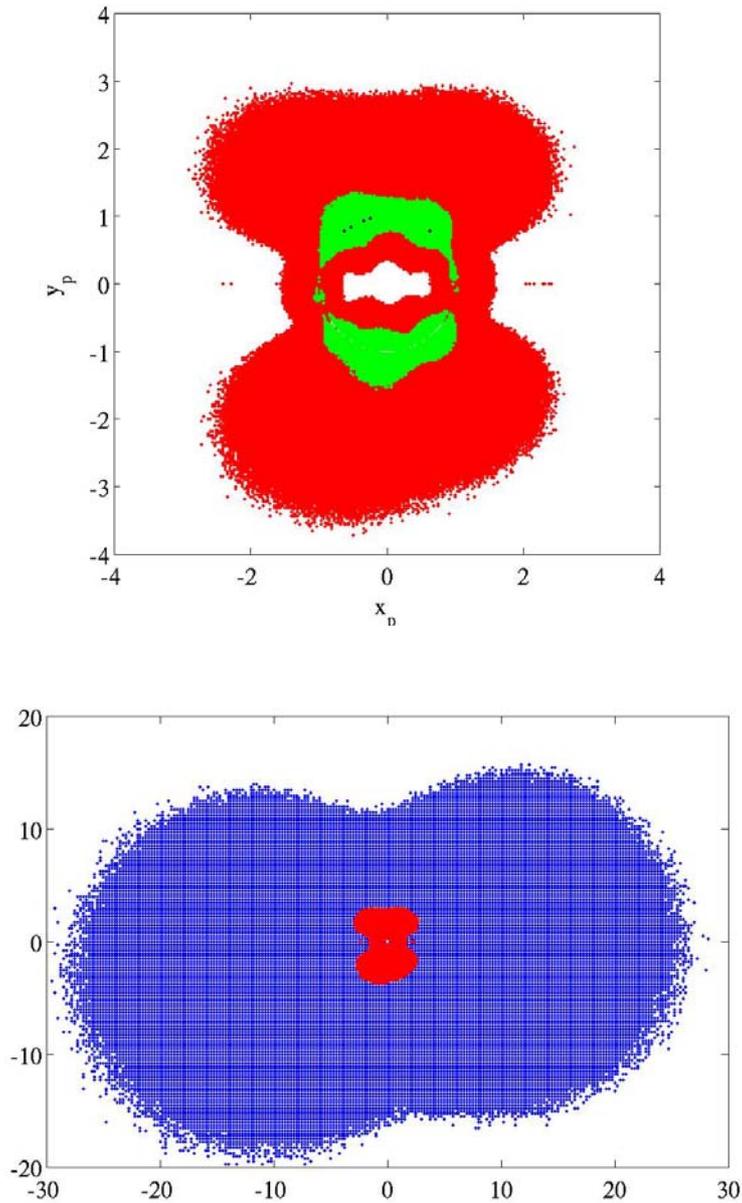

**Fig. 4**. Exclusion regions for planets orbiting the lens star of M32/O219 as a function of the planet:star mass ratio $q$, and the projected coordinates $x_p$ and $y_p$ of the planet in units of $r_E$. The source-size parameter $r_s/r_{E'} = 0.0007$. The upper panel shows exclusion regions for low-mass planets with $q = 10^{-6}$ (black), $10^{-5}$ (green) and $10^{-4}$ (red), respectively, and the lower panels shows exclusion regions for $q = 10^{-4}$ (red) and $10^{-3}$ (blue). The sizes of the exclusion regions do not depend critically on the peak magnification. The regions for the magnification at its 1-σ upper or lower limits would not be dissimilar (*18*). Indeed, doubling or halving the magnification does not dramatically alter the exclusion regions in these events (*10*). The orientation of the plots is similar to that used in (*5*), with the source star moving horizontally left to right, just beneath the lens star.

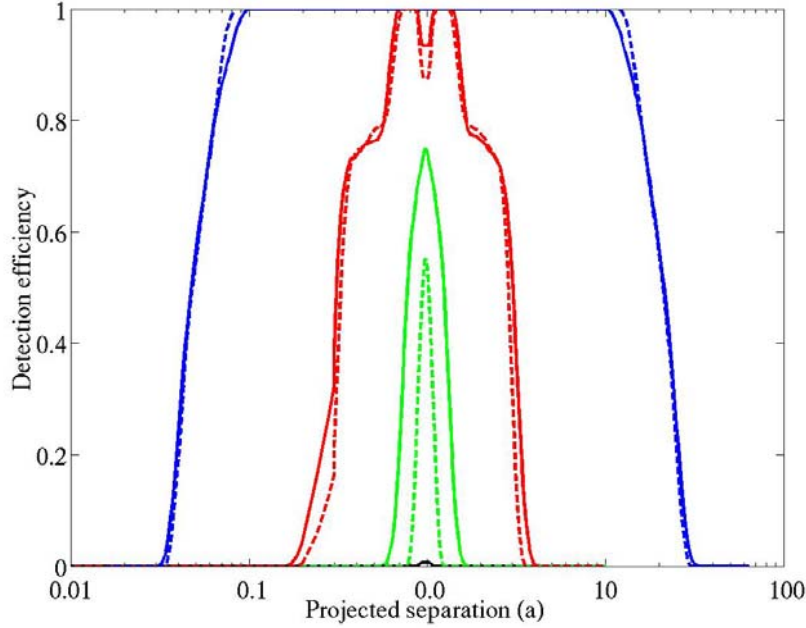

**Fig. 5.** Detection efficiencies for planets orbiting the lens star of M32/O219 as a function of the planet:star mass ratio $q$, and the projected separation $a$ in units of $r_E$. The $q$ values are $10^{-6}$ (black), $10^{-5}$ (green), $10^{-4}$ (red) and $10^{-3}$ (blue). The solid lines denote efficiencies for the source-size parameter $r_s/r_{E'} = 0.0007$, and the dashed lines for the (less likely) value of 0.0015. These results demonstrate the gradual loss of sensitivity that occurs for low mass planets as $r_s/r_{E'}$ increases. The results also illustrate a well-known degeneracy of the microlensing technique at high magnification. Planets at projected separations $a$ and $a^{-1}$ produce identical perturbations to the microlensing light curve (*2*). Information from space-based searches for extrasolar planets using the transit technique, which will provide statistics on the abundances of all types of planets at orbital separations < 1 AU, could assist to break this degeneracy. We note that the caustic-crossing technique (*6-9*) is not subject to the above degeneracy.